\documentstyle[amssymb,amsmath,prd,aps,preprint]{revtex}

\newcommand{\nc}{\newcommand}
\nc{\renc}{\renewcommand}
\nc{\be}[1]{\begin{equation} \mbox{$\label{#1}$}}
\nc{\bea}[1]{\begin{eqnarray} \mbox{$\label{#1}$}}
\nc{\Section}[2]{\section{#2}\label{#1}}
\nc{\Bibitem}[1]{\bibitem{#1}}
\nc{\Label}[1]{\label{#1}}
\nc{\eea}{\end{eqnarray}}
\nc{\ee}{\vspace{\undereqskip}\end{equation}}

\def\itt{}

\nc{\aap}[3]{\itt{  Astron.\ Astrophys.\ }{{\bf #1} {(#2)} {#3}}}
\nc{\advp}[3]{\itt{  Adv.\ in\ Phys.\ }{{\bf #1} {(#2)} {#3}}}
\nc{\annp}[3]{\itt{  Ann.\ Phys.\ (N.Y.)\ }{{\bf #1} {(#2)} {#3}}}
\nc{\apjl}[3]{\itt{  Ap.\ J.\ Lett.\ }{{\bf #1} {(#2)} {#3}}}
\nc{\app}[3]{\itt{ Astropart.\ Phys.\ }{{\bf #1} {(#2)} {#3}}}
\nc{\cmp}[3]{\itt{  Comm.\ Math.\ Phys.\ }{{ \bf #1} {(#2)} {#3}}}
\nc{\cqg}[3]{\itt{  Class.\ Quant.\ Grav.\ }{{\bf #1} {(#2)} {#3}}}
\nc{\epj}[3]{{ Eur.\ Phys.\ J.\ }{{\bf #1}, {#3} (#2)}}
\nc{\epl}[3]{\itt{  Europhys.\ Lett.\ }{{\bf #1},  {#3}  {(#2)}}}
\nc{\ijmp}[3]{\itt{ Int.\ J.\ Mod.\ Phys.\ }{{\bf #1},  {#3}  {(#2)}}}
\nc{\ijtp}[3]{\itt{ Int.\ J.\ Theor.\ Phys.\ }{{\bf #1},  {#3}  {(#2)}}}
\nc{\jmp}[3]{\itt{  J.\ Math.\ Phys.\ }{{ \bf #1} {(#2)} {#3}}}
\nc{\jpa}[3]{\itt{  J.\ Phys.\ A\ }{{\bf #1},  {#3}  {(#2)}}}
\nc{\jpc}[3]{\itt{  J.\ Phys.\ C\ }{{\bf #1},  {#3}  {(#2)}}}
\nc{\jap}[3]{\itt{ J.\ Appl.\ Phys.\ }{{\bf #1},  {#3}  {(#2)}}}
\nc{\jpsj}[3]{\itt{ J.\ Phys.\ Soc.\ Japan\ }{{\bf #1},  {#3}  {(#2)}}}
\nc{\lmp}[3]{\itt{ Lett.\ Math.\ Phys.\ }{{\bf #1},  {#3}  {(#2)}}}
\nc{\mpl}[3]{\itt{  Mod.\ Phys.\ Lett.\ }{{\bf #1},  {#3}  {(#2)}}}
\nc{\ncim}[3]{\itt{  Nuov.\ Cim.\ }{{\bf #1},  {#3}  {(#2)}}}
\nc{\npbb}[3]{\itt{  Nucl.\ Phys.\ B\ }{{\bf #1},  {#3}  {(#2)}}}
\nc{\prr}[3]{\itt{ Phys.\ Rev.\ }{{\bf #1},  {#3}  {(#2)}}}
\nc{\praa}[3]{\itt{  Phys.\ Rev.\ A\ }{{\bf #1},  {#3}  {(#2)}}}
\nc{\prbb}[3]{\itt{  Phys.\ Rev.\ B\ }{{{\bf #1},  {#3}  {(#2)}}}}
\nc{\prcc}[3]{\itt{  Phys.\ Rev.\ C\ }{{\bf #1},  {#3}  {(#2)}}}
\nc{\prdd}[3]{\itt{  Phys.\ Rev.\ D\ }{{\bf #1},  {#3}  {(#2)}}}
\nc{\prll}[3]{\itt{ Phys.\ Rev.\ Lett.\ }{{\bf #1},  {#3}  {(#2)}}}
\nc{\pll}[3]{\itt{  Phys.\ Lett.\ }{{\bf #1},  {#3}  {(#2)}}}
 \nc{\plbb}[3]{\itt{  Phys.\ Lett.\ B\ }{{\bf #1},  {#3}  {(#2)}}}
\nc{\prep}[3]{\itt{ Phys.\ Rep.\ }{{\bf #1},  {#3}  {(#2)}}}
\nc{\prsl}[3]{\itt{ Proc.\ R.\ Soc.\ London\ }{{\bf #1},  {#3}  {(#2)}}}
\nc{\ptp}[3]{\itt{  Prog.\ Theor.\ Phys.\ }{{\bf #1},  {#3}  {(#2)}}}
\nc{\ptps}[3]{\itt{  Prog\ Theor.\ Phys.\ suppl.\ }{{\bf #1},  {#3}  {(#2)}}}
\nc{\physa}[3]{\itt{  Physica\ A\ }{{\bf #1},  {#3}  {(#2)}}}
\nc{\physb}[3]{\itt{  Physica\ B\ }{{\bf #1},  {#3}  {(#2)}}}
\nc{\phys}[3]{\itt{ Physica\ }{{\bf #1},  {#3}  {(#2)}}}
\nc{\rpp}[3]{\itt{ Rep.\ Prog.\ Phys.\ }{{\bf #1},  {#3}  {(#2)}}}
\nc{\sjnp}[3]{\itt{ Sov.\ J.\ Nucl.\ Phys.\ }{{\bf #1},  {#3}  {(#2)}}}
\nc{\spjetp}[3]{\itt{ Sov.\ Phys.\ JETP\ }{{\bf #1},  {#3}  {(#2)}}}
\nc{\yf}[3]{\itt{ Yad.\ Fiz.\ }{{\bf #1},  {#3}  {(#2)}}}
\nc{\zetp}[3]{\itt{ Zh.\ Eksp.\ Teor.\ Fiz.\  }{{\bf #1}  {(#2)} {#3}}}
\nc{\zp}[3]{\itt{ Z.\ Phys.\ }{{\bf #1},  {#3}  {(#2)}}}
\nc{\ibid}[3]{{\sl ibid.\ }{{\bf #1},  {#3}  {(#2)}}}

\begin{document}

\title{ \vskip-5truecm{\hfill {\small CFNUL/01-05}} \\
\vskip-0.4truecm
{\hfill {\small DFAQ-2001/06-TH}} \\
Blocking Active-Sterile Neutrino Oscillations
 in the \\ Early Universe with a Majoron Field}
\author{Lu\'{\i}s Bento \\
{\normalsize \emph{Centro de F\'{\i}sica Nuclear da Universidade de Lisboa, }
} \\
\vskip-0.2truecm
{\normalsize \emph{Avenida Prof. Gama Pinto 2, 1649-003 Lisboa, Portugal }} 
\\ }
\author{
Zurab Berezhiani \\
{\normalsize \emph{
Dipartamento di Fisica, Universit\'a di L'Aquila, I-67010
Coppito, L'Aquila, Italy }} \\
\vskip-0.2truecm
{\normalsize \emph{ INFN, Laboratori Nazionali del Gran Sasso, 
I-67010 Assergi, L'Aquila, Italy }}\\
\vskip-0.2truecm
{\normalsize \emph{Andronikashvili Institute of Physics, GE-380077 Tbilisi,
Georgia }}}
\date{August, 2001}
\maketitle
\vskip-0.7truecm
\tightenlines
\begin{abstract}
We propose a new mechanism to block the active-sterile neutrino oscillations in the Early Universe.
We show that a typical consequence of theories where the lepton number is spontaneously broken is the existence of a coherent cosmological Majoron field with a strength proportional to the lepton and baryon numbers of the Universe.
This field interacts with leptons and changes the potentials relevant for neutrino oscillations.
If the scale of lepton number symmetry breaking is of the order of 1 GeV then a Majoron field and lepton number asymmetry of the order of the baryon asymmetry 
are strong enough to block the active-sterile neutrino oscillations with the atmospheric neutrino mass gap which otherwise would bring the sterile neutrino into equilibrium at the big bang nucleosynthesis epoch. 

\strut 

PACS numbers: 14.80.Mz, 14.60.Pq, 95.30.Cq, 98.80.Ft 
\end{abstract}


\newpage

\section{Introduction}

\label{introduction}

The explanation of the present neutrino puzzles~\cite{reviews} may require the existence of
one or several species of extra light sterile neutrinos~\cite{sterile}. 
In particular, the sterile neutrino could be relevant for the explanation of
the atmospheric neutrino problem (ANP)~\cite{skatm98} in the presence of a
significant $\nu _{\mu }-\nu _{s}$ mixing\footnote{
The recent Super-Kamiokande data can be explained by 
$\nu _{\mu }-\nu _{\tau }$ oscillations
while the situation where the ANP is exclusively due to 
$\nu _{\mu }-\nu_{s}$ 
oscillations is disfavored~\cite{skatm00}. However, the more general
case where $\nu _{\mu }$ oscillates into $\nu _{\tau }$ and $\nu _{s}$ with
comparable rates is completely consistent with the data~\cite{foot00}.}.
The typical required values are $\delta m_{\mathrm{atm}}^{2}\sim 3\times 10^{-3}$
eV$^{2}$ and large mixing angle, $\sin ^{2}2\theta _{\mathrm{atm}}\simeq 1$.

On the other hand, for such a parameter range one can encounter a
contradiction with the Big Bang nucleosynthesis (BBN) 
bounds~\cite{BBN} on the number
of extra light particle species; namely, according to the analyses of 
ref.~\cite{dolgov81}, the sterile neutrino comes into equilibrium with the
particle thermal bath via $\nu _{\mu }-\nu _{s}$ oscillation, unless the
condition $\delta m^{2}\sin ^{4}2\theta \lesssim 3\times 10^{-6}$ eV$^{2}$
is satisfied 
(updated constraints are given in ref.~\cite{kirilova97} for small mass differences, $\delta m^{2} \lesssim  10^{-7}$ eV$^{2}$), 
which is certainly out of the range of parameters needed to
explain the ANP.

However, it was found~\cite{foot95} that the $\nu _{\mu }-\nu _{s}$
oscillations are suppressed at temperatures $T\lesssim 3$ MeV (the
decoupling temperature of $\nu _{\mu ,\tau }$) if the lepton number
asymmetry at these temperatures is very high, namely $L_{a}\gtrsim 10^{-5}$
(lepton number to photon number 
ratio). 
But this is $4-5$ orders of
magnitude larger than the observed baryon asymmetry of the universe ($
B\lesssim 10^{-9}$) and in the most generic baryogenesis context one can
expect that $L\sim B$ (e.g., in the context of grand unified theory
(GUT) baryogenesis or
leptogenesis~\cite{fukujita86,buchmuller99} this is because the $B+L$
non-conserving sphaleron processes redistribute $B$ and $L$ among each other~
\cite{kuzmin85,khlebnikov88,harvey90}). The same is true in the context of
the electroweak baryogenesis ($B-L=0$). In the Affleck-Dine mechanism $B$
and $L$ can in principle be independent of each other, but still of the
same order.

It has been shown~\cite{foot96,kirilova97} that at much lower temperatures ($
T<100$ GeV) neutrino oscillations can actually produce a rapid increase of
the lepton asymmetries from the initial very small values up to the order of 
$0.1$. This however only occurs for negative $\delta m^{2}\cos 2\theta $ and
very small active-sterile mixing angles, not directly relevant for the ANP.

In this paper we show that a lepton asymmetry as small as the present baryon
asymmetry may be enough to block the sterile neutrino oscillations. The
necessary new ingredient is the existence of a coherent Majoron
 field in the Early Universe and a low scale of spontaneous
breaking of lepton number, $F_{L}\sim 1$ GeV
(The Majoron~\cite{chikashige81} is the massless Nambu-Goldstone boson in models where the total or any partial lepton number is spontaneously broken.).

While it has been common wisdom that due to their derivative coupling
nature~\cite{gelmini83} Nambu-Goldstone bosons cannot mediate long range
interactions, it has recently been demonstrated~\cite{bento98} that a
coherent source of a Majoron field, to be specific, is formed whenever the
corresponding broken lepton number suffers a net increase or decrease in a
certain region of space. The processes that violate this lepton number can
be the very neutrino oscillations as exemplified in previous papers~\cite
{bento98} or any other reactions. In the present work we show that a Majoron
field can be produced due to lepto- and baryogenesis processes in the Early
Universe. The Majoron field interacts with neutrinos with a strength
inversely proportional to the lepton breaking scale, $F_{L}$. If $F_{L}$ is
around $1$ GeV, a lepton asymmetry as small as $L\sim B\sim 10^{-9}$ can
block the $\nu _{\mu }$ oscillation into sterile neutrinos with $\delta
m^{2}\sim 3\times 10^{-3}$\textrm{\ eV}$^{2}$ no matter how large their
mixing angle is. That is our thesis.

The origin of the Majoron field is elaborated in section \ref{majoron} and
its role on neutrino oscillations into a sterile neutrino in section \ref
{oscillations}. But first we build in section \ref{model} a specific model
of neutrino masses with spontaneous breaking of lepton number and in
section \ref{asymmetries} we derive the relations between the particle
asymmetries in the Early Universe and the present baryon number. The
model aims to fit the present known observations from solar,
atmospheric and terrestrial neutrino experiments, 
including the Liquid Scintillation Neutrino Detector
(LSND) result~\cite{LSND},
with an extra sterile neutrino. However, it should be emphasized that the
mechanism we propose of suppression of oscillations into a sterile neutrino
at the BBN epoch, based on the existence of a Majoron field, does not depend
on the particular model or set of neutrino mass parameters. The only
fundamental assumptions are that the lepton number (or a partial lepton
number) is spontaneously broken and the breaking scale is around the 1 GeV
magnitude.
In the next section we also make the point that in the absence of the LSND neutrino mass gap the oscillations of atmospheric and solar neutrinos into a sterile neutrino are no longer correlated with each other.
In the last section we draw our conclusions.


\section{Neutrino masses and mixing}
\label{mixing}

The existence of a fourth, sterile, neutrino has been suggested
as it is the only way of reconciling the atmospheric, solar,
and\ LSND neutrino oscillation evidence and their very different $\delta
m^{2} $ mass gap scales. The wide mass gap (${\mathcal{O}}(1\mathrm{\ eV})$)
that is necessary to explain the LSND result in terms of $\nu _{e}-\nu _{\mu
}$ mixing requires that the neutrino mass pattern should have a two-doublet
structure~\cite{bilenky98}: one of the doublets consists of $\nu _{e}$ and $
\nu _{s}$ or $\nu _{\tau }$, or a linear combination of both, and is
responsible for the solar neutrino deficit, and the other one, responsible
for the atmospheric neutrino anomaly, consists of $\nu _{\mu }$ and $\nu
_{\tau }$ (or a linear combination of $\nu _{s}$ and $\nu _{\tau }$). These
doublets are separated by the LSND mass gap.

It has been shown~\cite{bahcall01,barger01} that even after the recent SNO
observations~\cite{SNO01} both the sterile and active neutrino oscillations
are viable solutions of the solar neutrino problem as well as a more general
superposition of both. On the other hand, the atmospheric neutrino data
seem~\cite{skatm00} to favor the $\nu _{\tau }$ solution against the sterile neutrino case
but the analysis~\cite{foot00} of the most recent data still allows a quite large relative
probability, more than 50\%, of oscillation into $\nu _{s}$ (the larger the
probability, $\sin ^{2}\xi $, the smaller the allowed $\delta m^{2}$ range).

A consequence of the LSND large mass gap and the limits from reactor
disappearance experiments such as Chooz~\cite{CHOOZ} is that the
solar neutrinos $\nu _{e}$ and the atmospheric neutrinos $\nu _{\mu }$ must oscillate
into states that are essentially orthogonal to each other. In other words,
if the solar electron neutrinos oscillate into the linear combination $\nu _{
\bar{e}}\equiv \cos \xi \;\nu _{s}-\sin \xi \;\nu _{\tau }$, then, the
atmospheric muon neutrinos necessarily oscillate into the state $\nu _{\bar{
\mu}}\equiv \sin \xi \;\nu _{s}+\cos \xi \;\nu _{\tau }$. 
However, the situation would be totally different if the LSND evidence was not present.

To be more specific, let $\nu _{1}$, $\nu _{2}$ and $\theta _{\odot }$\ be
the mass eigenstates and mixing angle responsible for the solar neutrino
deficit and $\nu _{3}$, $\nu _{4}$ and $\theta _{\mathrm{atm}}$\ the states
and mixing angle relevant for atmospheric neutrinos. 
The two pairs are
separated by the LSND mass gap and no other specific mass hierarchy has to
be assumed. The reactor experiments constrain the mixing matrix elements $
U_{e3}$, $U_{e4}$, $U_{\mu 1}$, and $U_{\mu 2}$ to be small~\cite{bilenky98}, but
not $U_{\tau i}$ or $U_{si}$. If one neglects all the mixing angles that are
necessarily small and irrelevant to explain the present bulk of data, the
mixing matrix is given as 
\begin{subequations} \label{4nmlsnd}
\begin{eqnarray}
\nu _{1}& =& \cos \theta _{\odot }\;\nu _{e}-\sin \theta _{\odot }(\cos \xi
\;\nu _{s}-\sin \xi \;\nu _{\tau })\;,  \\
\nu _{2}& =& \sin \theta _{\odot }\;\nu _{e}\;+\cos \theta _{\odot }(\cos \xi
\;\nu _{s}-\sin \xi \;\nu _{\tau })\;,   \\
\nu _{3}& =& \cos \theta _{\mathrm{atm}}\;\nu _{\mu }-\sin \theta _{\mathrm{atm
}}(\sin \xi \;\nu _{s}+\cos \xi \;\nu _{\tau })\;,   \\
\nu _{4}& =& \sin \theta _{\mathrm{atm}}\;\nu _{\mu }+\cos \theta _{\mathrm{atm
}}(\sin \xi \;\nu _{s}+\cos \xi \;\nu _{\tau })\;.
\end{eqnarray}
\end{subequations}
The mass eigenstates $\nu _{1}$ and $\nu _{2}$ are separated by the gap $
\delta m_{\odot }^{2}$ and $\nu _{3}$, $\nu _{4}$ by $\delta m_{\mathrm{atm}
}^{2}$.

Clearly, the less the atmospheric neutrinos oscillate into the sterile
neutrino the more the solar neutrinos have to oscillate into $\nu _{s}$.
There is a potential clash in the future if both solar and atmospheric
neutrino experiments happen to constrain the respective sterile
neutrino solutions to less than a 50\% probability. In that case the conflict
with the LSND data will be insoluble, which will call for new results from
MiniBooNE~\cite{boone00}, the next new independent accelerator experiment. 
Suppose for a moment
that the LSND evidence does not exist or is going to be ruled out by the
MiniBooNE experiment. We would like to stress that this does not rule out the
sterile neutrino as a possible protagonist in the other solar and
atmospheric neutrino problems, not even if both of them exclude dominant
sterile neutrino solutions. On the contrary, the absence of the LSND mass
gap increases the freedom in the neutrino mixing parameters. 
Then, the two-doublet mass pattern is no longer inevitable and 
the role of $\nu _{s}$ in
the solar neutrino deficit is completely decoupled from its role in the
atmospheric neutrino oscillations.

As a matter of proof we make explicit an extreme case, namely, where the
atmospheric neutrinos oscillate into $\nu _{\tau }$ or $\nu _{s}$ with
arbitrary relative probabilities, while the solar neutrinos oscillate
exclusively to $\nu _{\tau }$ and $\nu _{\mu }$ but not to $\nu _{s}$. The
mixing matrix can be described as follows: 
\begin{subequations} \label{4nm}
\begin{eqnarray}
\nu _{1}& =& \cos \theta _{\odot }\;\nu _{e}-\sin \theta _{\odot }(-\sin
\alpha \;\nu _{\mu }+\cos \alpha \;\nu _{\tau })\;,   \\
\nu _{2}& =& \sin \theta _{\odot }\;\nu _{e}\;+\cos \theta _{\odot }(-\sin
\alpha \;\nu _{\mu }+\cos \alpha \;\nu _{\tau })\;,  \\
\nu _{3}& =& \sin \beta \;\nu _{s}+\cos \beta \,(\cos \alpha \;\nu _{\mu
}+\sin \alpha \;\nu _{\tau })
\;,   \\
\nu _{4}& =& \cos \beta \;\nu _{s}-\sin \beta \,(\cos \alpha \;\nu _{\mu
}+\sin \alpha \;\nu _{\tau })
\;.
\end{eqnarray}
\end{subequations}
As far as
the mass spectrum is concerned, the mass eigenstate $\nu _{3}$ is separated
from the other three by mass gaps that are in the atmospheric neutrino range $\sim
\delta m_{\mathrm{atm}}^{2}\sim 3\times 10^{-3}$ eV$^{2}$.
 $\nu _{1}$ and $
\nu _{2}$ are almost degenerate and separated by the solar neutrino mass
gap and finally, $\nu _{4}$\ is only subject to the condition $
m_{3}^{2}-m_{4}^{2}\sim \delta m_{\mathrm{atm}}^{2}$, as it, like $\nu _{3}$, does not participate in the solar neutrino oscillations.
The mixing angles relate to the atmospheric mixing angle as $\cos \theta _{
\mathrm{atm}}=\cos \alpha \cos \beta $. The atmospheric neutrinos $\nu _{\mu
}$ oscillate into$\;\nu _{\tau }$ with a probability proportional to $\sin
^{2} \alpha \cos ^{2}\beta $ whereas the probability of oscillation into
$\;\nu _{s}$ is proportional to $\sin ^{2}\beta $. The ratio between them
is given by $\tan ^{2}\xi =\tan ^{2}\beta /\sin ^{2}\alpha $. It is
clear that the solar neutrinos do not oscillate into $\nu _{s}$.

This just shows that the potential problem raised by the possibility that the
atmospheric neutrinos oscillate significantly into a sterile neutrino with
its consequences for BBN is not necessarily linked to the solar neutrino
solutions and does not depend on the LSND evidence although it has been
motivated by the coexistence of all three kinds of observation. In the
present work we want to present a solution and a mechanism to block the
oscillations of muon neutrinos into sterile neutrinos in the Early Universe
at the time of BBN. The idea does not depend crucially on the particular
neutrino mixing pattern but the actual numbers vary, of course, from model to
model. We worked out in detail a particular model that is suitable to
encompass all three types of neutrino oscillation evidence, including LSND.


\section{Neutrino mass model}
\label{model}

The seesaw mechanism~\cite{seesaw} can be incorporated within a model where the lepton
number is spontaneously broken at a relatively low energy scale by adding to the standard lepton doublets 
$\ell_{i}$ and charged singlets $e_{i}$
two heavy sterile neutrinos per lepton generation, $N_{Li}$ and $N_{Ri}^{C}$
 (left-handed), with lepton
numbers $+1$ and $-1$, respectively. 
The additional light sterile neutrino, $\nu _{s}$ (left-handed),
 has lepton number $L_{s}=-3$. 
 The most
general Yukawa interaction Lagrangian in the lepton sector is written in
Majorana matrix form as 
\begin{equation}
{\mathcal{L}}_{Y}=\frac{1}{2}\, \psi ^{T}C{\mathcal{M}}\psi +\mathrm{H.C.}
\;,  \label{ly}
\end{equation}
where $\psi \equiv (e_{i}^{C},\ell _{i},\nu _{s},N_{L\,i},N_{R\,i}^{C})$ and 
${\mathcal{M}}$ is the symmetric matrix

\begin{equation}
{\mathcal{M}}= 
\begin{tabular}{c|ccccc}
& $e_{j}^{C}$ & $\ell_{j}$ & $\nu_{s}$ & $N_{L\,j}$ & $N_{R\,j}^{C}$ \\ 
\hline
$e_{i}^{C}$ & $0$ & $h_{e}^{T}H_{1}$ & $0$ & $0$ & $0$ \\ 
$\ell_{i}$ & $-$ & $0$ & $0$ & $0$ & $h_{N}H_{2}$ \\ 
$\nu_{s}$ & $-$ & $-$ & $0$ & $h_{s}^{T}\sigma^{\ast}$ & $0$ \\ 
$N_{L\,i}$ & $-$ & $-$ & $-$ & $h_{L}\sigma$ & $M$ \\ 
$N_{R\,i}^{C}$ & $-$ & $-$ & $-$ & $-$ & $h_{R}\sigma^{\ast}$
\end{tabular}
\mathrm{\;}.  \label{m}
\end{equation}
The omitted elements are obtained by symmetrization. $H_{1}$ and $H_{2}$ are
two standard Higgs doublets under SU(2) and $\sigma$ is the singlet scalar
field with lepton number $L_{\sigma}=-2$. $h_{s}$ is a $3\times1$ column and 
$h_{e}$, $h_{N}$, $h_{L}$, $h_{R}$, and $M$ are $3\times3$ matrices.

Before lepton number spontaneous breaking the heavy sterile neutrinos form
Dirac particles, namely, $N_{i}=N_{L\,i}+N_{R\,i}$, with lepton number equal to 1
and masses $M_{i}$ in the basis where $M$ is diagonal: 
$M={\mathrm{diag}}(M_{i})$. 
After lepton and gauge symmetry breaking the light neutrinos
acquire masses and mix with the sterile neutrino in a $4\times 4$ Majorana
mass matrix. Denoting the $3\times 3$ active, $3\times 1$ active-sterile and 
$1\times 1$ sterile neutrino blocks respectively, as $m_{\nu \nu }$, $m_{\nu
s}$ and $m_{ss}$, we obtain in leading order in any basis where $M$ is a real
matrix 
\begin{subequations} \label{nmm}
\begin{eqnarray}
m_{\nu \nu }& =& h_{N}M^{-1}h_{L}(h_{N}M^{-1})^{T}\,\langle \sigma \rangle
\,v_{2}^{2}\;,  \label{mnn} \\
m_{\nu s}& =& -h_{N}M^{-1}h_{s}\,\langle \sigma \rangle \,v_{2}\;,\,
\label{mns} \\
m_{ss}& =&(M^{-1}h_{s})^{T}h_{R}M^{-1}h_{s}\,\langle \sigma \rangle ^{3}\;,
\label{mss}
\end{eqnarray}
\end{subequations}
where $v_{2}=\langle H_{2}^{0}\rangle $. We take as reference scales $m_{\nu \nu }\sim 0.05$ eV to account
for the atmospheric neutrino anomaly and $m_{\nu s}\sim 1$ eV for the LSND $
\nu _{\mu }(\bar{\nu}_{\mu })\rightarrow \nu _{e}(\bar{\nu}_{e})$ evidence.
Since we also assume $\langle \sigma \rangle \sim 1$ GeV the element $
m_{ss}\sim 10^{-13}$ eV is completely negligible.


\section{Asymmetries in the Early Universe}

\label{asymmetries}

At temperatures below the heavy neutrino masses $M_{i}\gtrsim10^{6}$ GeV, the
Dirac masses $M_{i}$ still mediate scattering processes capable of producing the
light singlet particles $\nu_{s}$ and $\sigma$ like $\ell H_{2}\rightarrow\bar{\nu}_{s}\sigma$. 
They can be studied in terms of the effective operators 
\begin{equation}
{\mathcal{L}}_{\mathrm{eff}}=\ell_{i}H_{2}\frac{m_{ij}}{2\langle\sigma\rangle
v_{2}^{2}}\ell_{j}H_{2}\,\sigma+\ell_{i}H_{2}\frac{m_{is}}{\langle
\sigma\rangle v_{2}}\nu_{s}\,\sigma^{\ast}+\nu_{s}\frac{m_{ss}}{2\langle
\sigma\rangle^{3}v_{2}}\nu_{s}\,\sigma^{\ast3}+\mathrm{H.C.\;,}  \label{leff}
\end{equation}
which also give rise to the light neutrino masses after spontaneous breaking of lepton number. 
One obtains the c.m.\ cross sections of the scattering
processes 1) $\bar{\ell}_{i}\bar{\ell}_{j}\rightarrow H_{2}H_{2}\sigma$, 
2) $\sigma\bar{H}_{2}\rightarrow\ell_{i}\nu_{s}$,
and 3) $\sigma\sigma\rightarrow \bar{\sigma}\nu_{s}\nu_{s}$ as 
\begin{align}
\sigma_{1} & =\frac{6\,s}{(8\pi)^{3}}\frac{|m_{ij}|^{2}}{\langle
\sigma\rangle^{2}v_{2}^{4}}\approx\frac{T^{2}}{\langle\sigma
\rangle^{2}v_{2}^{4}}\times10^{-5}\mathrm{\;eV}^{2}\;, \\
\sigma_{2} & =\frac{1}{8\pi}\frac{|m_{is}|^{2}}{\langle\sigma\rangle
^{2}v_{2}^{2}}\approx\frac{4}{\langle\sigma\rangle^{2}v_{2}^{2}}\times
10^{-2}\mathrm{\;eV}^{2}\;, \\
\sigma_{3} & =\frac{6\,s}{(8\pi)^{3}}\frac{|m_{ss}|^{2}}{\langle
\sigma\rangle^{6}}\approx\frac{T^{2}}{\langle\sigma\rangle^{6}}\times
10^{-28}\mathrm{\;eV}^{2}\;,
\end{align}
respectively,
where we have summed over initial and final weak isospin states ($\sqrt{s}$
is the c.m.\ energy).

In each case one compares the rate of collisions per particle, $\Gamma
=\sigma \,n\approx 0.1\,\sigma \,T^{3}$ (the boson number density is $
n_{b}\simeq 0.122\,T^{3}$ and the fermion number density $n_{f}\simeq
0.091\,T^{3}$) with the Hubble rate $H\approx T^{2}/10^{18}$ GeV, assuming a
total number of degrees of freedom around 100. The scalar singlet $\sigma $
is produced through the processes 1) $\bar{\ell}_{i}\bar{\ell}
_{j}\rightarrow H_{2}H_{2}\sigma $, $\bar{\ell}_{i}\bar{H}_{2}\rightarrow
\ell _{j}H_{2}\sigma $, and $\bar{H}_{2}\bar{H}_{2}\rightarrow \ell _{i}\ell
_{j}\sigma $ with cross sections $\sigma _{1}$, $2\sigma _{1}/3$, and 
$\sigma _{1}/3$, respectively (for $i\neq j$), 
which gives a total rate per Hubble time 
$\Gamma _{\sigma }H^{-1}\approx 2(v/v_{2})^{4}T^{3}/10^{15}$\textrm{\ GeV}$
^{3}$ ($v\simeq 174$\textrm{\ GeV} is the electroweak breaking
scale). 
This shows that $\sigma $ is in thermal equilibrium at temperatures
larger than $T_{\sigma }\approx 10^{5}$\textrm{\ GeV}, 
if one takes $v_{2}=v$. 

The sterile neutrino is produced in the processes 2) 
$\sigma \bar{H}_{2}\rightarrow \ell _{i}\nu _{s}$, 
$\bar{\ell}_{i}\sigma \rightarrow
H_{2}\nu _{s}$, and $\bar{\ell}_{i}\bar{H}_{2}\rightarrow \bar{\sigma}\nu
_{s}$, with cross sections $\sigma _{2}$, $\sigma _{2}/2$, and 
$\sigma _{2}/2$, respectively, and a total rate 
$\Gamma _{s}H^{-1}\approx
(v/v_{2})^{2}T/T_{s}$, which makes the decoupling temperature of
the light sterile neutrino $T_{s}\approx 4\times 10^{6}$\textrm{\ GeV}. 
If one or more heavy neutrinos $N_{i}$ have masses under that value, 
$\nu_{s}$ may decouple when some of the $N_{i}$ degrees of freedom 
are still present in the Universe 
(note that $M_{i}\gtrsim 10^{6}$\textrm{\ GeV}). 
Finally, processes like $\sigma \sigma
\rightarrow \bar{\sigma}\nu _{s}\nu _{s}$ are too weak to be relevant.

Above $T_{s}$ the sterile neutrino and scalar singlet $\sigma $ are in
chemical equilibrium with the lepton and Higgs doublets and their number
asymmetries are constrained by the equations of detailed balance. The
precise relations between the particle asymmetries depend on which particles
and processes are in thermodynamical equilibrium at a given time. To be
definite we assume that by the time the sterile neutrino decouples, $B$ and $
L$ are only violated by electroweak instanton processes while $B-L$ is
conserved. On the other hand the right-handed electrons $e_{R}$ are not yet
in chemical equilibrium and the quarks $u_{R}$, $d_{R}$ may or may not be
in equilibrium depending on the exact values of their Yukawa couplings and
temperature $T_{s}$. In either case the equations of detailed balance yield
the particle asymmetries as functions of the $B-L$ asymmetry.

At temperatures above $T_{s}$ the operators of eq.~(\ref{leff}) yield the
chemical potential constraints 
\begin{subequations} \label{nedb}
\begin{eqnarray}
\mu _{\sigma }+2\mu _{\ell }+2\mu _{H}& =& 0\;,  \label{musigma} \\
\mu _{s}-\mu _{\sigma }+\mu _{\ell }+\mu _{H}& =& 0\;.  \label{mus}
\end{eqnarray}
\end{subequations} 
The other constraints come from standard model 
reactions~\cite{khlebnikov88,harvey90}. 
To be definite we assume that $u_{R}$ and $d_{R}$
are in equilibrium at $T_{s}$ (the temperature at which they come into
equilibrium increases with increasing Yukawa couplings and therefore with 
increasing number of Higgs doublets). 
The electroweak and QCD instantons and the Yukawa
interactions imply that 
\begin{subequations} \label{edb}
\begin{eqnarray}
3\mu _{q}+\mu _{\ell }& =& 0\;,  \label{ewinstantons} \\
2\mu _{q}-\mu _{u}-\mu _{d}& =& 0\;,  \label{qcdinstantons} \\
\mu _{q}-\mu _{d}-\mu _{H}& =& 0\;,  \label{dyukawa} \\
\mu _{q}-\mu _{u}+\mu _{H}& =& 0\;,  \label{uyukawa} \\
\mu _{\ell }-\mu _{\tau }-\mu _{H}& =& 0\;,  \label{lyukawa} 
\end{eqnarray}
\end{subequations}
where $\mu _{q}$, $\mu _{\ell }$, and $\mu _{H}$ designate the flavor
universal chemical potentials of the quark, lepton, and Higgs doublets,
respectively, 
$\mu _{u}$, $\mu _{d}$ those of the right-handed quark isosinglets
and $\mu _{\tau }$ the common chemical potential of the lepton isosinglets 
$\mu _{R}$ and $\tau _{R}$. Since the electron singlet $e_{R}$ is not in
chemical equilibrium its chemical potential $\mu _{e}$ is an independent
variable. We may assume that a baryon asymmetry originally produced in a GUT
baryogenesis scenario is later communicated through electroweak instantons
to the lepton sector ($T\lesssim 10^{12}$\textrm{\ GeV}) but not to $e_{R}$.
In that case $\mu _{e}$ remains zero until the $e_{R}$ Yukawa interactions
come into equilibrium at temperatures lower than $T_{s}$.

A vanishing weak hypercharge implies 
\begin{equation}
3(\mu _{q}+2\mu _{u}-\mu _{d}-\mu _{\ell })-2\mu _{\tau }-\mu _{e}+2n_{H}\mu
_{H}=0\;.  \label{hypercharge}
\end{equation}
Here $n_{H}$ is the total number of Higgs doublets; $n_{H}=1$ if $H_{1}$ and 
$H_{2}$ are the same field and $n_{H}=2$ otherwise. The above constraints
and the condition $\mu _{e}=0$ leave only one independent variable. It is
convenient to choose this as $\mu _{s}$\ because the $\nu _{s}$ abundance
and number asymmetry are conserved after its decoupling. The other quantity
that is conserved is $B-L$. Denoting the baryon number density as $dB/dV= 
\bar{B}T^{3}/6$ and likewise for $L$ and $B-L$, one has 
\begin{subequations} \label{BLLsnot}
\begin{eqnarray}
\bar{B}& =& 6\mu _{q}+3\mu _{u}+\mu _{d}\;,  \label{B} \\
\bar{L}& =& \bar{L}_{/\hspace{-0.39pc}s}-3\mu _{s}\;,  \label{L} \\
\bar{L}_{/\hspace{-0.39pc}s}& =& 6\mu _{\ell }+2\mu _{\tau }+\mu _{e}-4\mu
_{\sigma }+2n_{N}\mu _{N}\;,  \label{Lnots}
\end{eqnarray}
\end{subequations} 
where $L_{/\hspace{-0.39pc}s}$ stands for the lepton number of all particles
except $\nu _{s}$ and $n_{N}$ is the number of relativistic Dirac heavy
neutrinos at a given moment. The decay processes $N_{i}\rightarrow \ell
_{j}H_{2}$ set the equation $\mu _{N}=\mu _{\ell }+\mu _{H}$. 
Putting everything together, one obtains 
\begin{align}
B-L_{/\hspace{-0.3915pc}s}& =\frac{1}{3}\left( 8+2n_{N}-\frac{15+4n_{H}}{
9+n_{H}}\right) (N_{s}-N_{\bar{s}})\;,  \label{B-Lnots} \\
\frac{B-L}{B-L_{/\hspace{-0.39pc}s}}& =\frac{(9+n_{H})(17+2n_{N})-15-4n_{H}}{
(9+n_{H})(8+2n_{N})-15-4n_{H}}\;.  \label{B-L/B-Lnots}
\end{align}
After $\nu _{s}$ decoupling, $B-L$, $B-L_{/\hspace{-0.39pc}s}$, and the 
$\nu _{s}$ number asymmetry $N_{s}-N_{\bar{s}}$ are all conserved. Although
the above relations are strictly valid only when $\nu _{s}$ is in thermal
equilibrium, one expects that the decoupling process does not introduce very
large perturbations and one may use these results as a first approximation.
They give the $\nu _{s}$ number asymmetry and $B-L_{/\hspace{-0.39pc}s}$ as
functions of the primordial $B-L$ and number $n_{N}$ of heavy neutrinos that
are relativistic when $\nu _{s}$ decouples.

The next transition is the decoupling of the scalar singlet $\sigma $ at a
temperature $T_{\sigma }$ around $10^{5}$\textrm{\ GeV}. This is
close to the $e_{R}$ coupling epoch which starts at $T_{e}\sim
(v/v_{1})^{2}\times 10^{4}$\textrm{\ GeV}. 
This temperature rises with increasing
electron Yukawa coupling and if one assumes the existence of two Higgs
doublets and $v_{1}\lesssim v/3$ ($v\simeq 174$\textrm{\ GeV}) then
 $e_{R}$ is already in equilibrium when $\sigma $ decouples. 
To be definite we assume so. 
One repeats the exercise with Eqs.\ (\ref{edb}) and (\ref
{hypercharge}), complemented with $\mu _{\tau }=\mu _{e}$ and Eq.~(\ref
{musigma}), to obtain all chemical potentials in terms of $\mu _{\sigma }$. 
Then Eqs.~(\ref{B}) and (\ref{Lnots}) with $n_{N}=0$ yield the relation 
between the $\sigma $ number asymmetry and the baryon and lepton numbers. 
Denoting the
total lepton number of the standard model particles as $L_{\ell }$ ($L_{\ell
}=L_{e}+L_{\mu }+L_{\tau }$) one derives for $n_{H}=2$, 
\begin{align}
N_{\sigma }-N_{\bar{\sigma}}& =\frac{12}{23}(B-L_{\ell })\;,  \label{nsigma}
\\
N_{s}-N_{\bar{s}}& =\frac{47}{23}\frac{33}{65+22n_{N}}(B-L_{\ell })\;,
\label{ns} \\
\frac{B-L}{B-L_{\ell }}& =\frac{47}{23}\frac{164+22n_{N}}{65+22n_{N}}\;,
\label{B-L/B-Ll}
\end{align}
where in the last two equations we used Eqs.~(\ref{B-Lnots}) and (\ref
{B-L/B-Lnots}), keeping in mind that $n_{N}$ is the number of Dirac neutrinos 
$N_{i}$ that are relativistic when $\nu _{s}$ decouples. Again, one expects
that the above results remain a reasonable approximation when $\sigma $
decouples. From then on, $B-L_{\ell }$, $B-L$, and the $\sigma $ and $\nu
_{s}$ abundances are conserved by all effective interactions.

When the temperature drops down to the electroweak phase transition the weak
isospin and hypercharge are no longer conserved, contrary to the electric
charge. The quarks and charged leptons form Dirac mass eigenstates whose
well defined chemical potentials are subject to a new set of 
constraints~\cite{harvey90}, together with the neutrinos and charged Higgs and 
$W$ bosons as follows: 
\begin{eqnarray}
\mu _{W^{+}}=\mu _{u}-\mu _{d}=\mu _{\nu }-\mu _{e}=\mu _{H^{+}}\;, \\
3\mu _{u}+3\mu _{d}+\mu _{\nu }+\mu _{e}=0\;.
\end{eqnarray}
The latter constraint is due to sphaleron processes. 
On the other hand, the net electric charge is zero: 
\begin{equation}
3(4\mu _{u}-2\mu _{d}-2\mu _{e})+2(2+n_{H})\mu _{W^{+}}=0\;.
\end{equation}
Definite $B$ and $L$ numbers are predicted in terms of the preexisting 
$B-L_{\ell }$ number, namely~\cite{harvey90}, 
\begin{align}
B& =\frac{32+4n_{H}}{98+13n_{H}}(B-L_{\ell })\;, \\
L_{\ell }& =-\frac{66+9n_{H}}{98+13n_{H}}(B-L_{\ell })\;.
\end{align}
These relations are preserved during the phase transition if there is no
intrinsic electroweak baryogenesis. After that the sphaleron processes stop
being effective and the baryon number is separately conserved. This allows us
to predict the scalar $\sigma $ particle and sterile neutrino asymmetries in
terms of the present baryon number. From Eqs.~(\ref{nsigma}) and (\ref{ns}),
valid for two Higgs doublets ($n_{H}=2$), one derives the asymmetries and
lepton numbers $L_{(\sigma )}=-2(N_{\sigma }-N_{\bar{\sigma}})$\ and $
L_{(s)}=-3(N_{s}-N_{\bar{s}})$ carried by $\sigma $ and $\nu _{s}$ as 
\begin{align}
L_{(\sigma )}& =-A_{\sigma }B=-\frac{24}{23}\frac{31}{10}B\;,  \label{lsigma}
\\
L_{(s)}& =-\frac{47}{23}\frac{31}{10}\frac{99}{65+22n_{N}}B\;.  \label{ls}
\end{align}
It is important to notice that, after spontaneous breaking of lepton
number, the $B-L$ violating processes are too weak to be in equilibrium, in
particular during the electroweak phase transition if it occurs after $L$
breaking. If that was not the case the baryon and lepton numbers would be
washed out. As soon as the sphalerons decouple the baryon and lepton numbers
start to be separately conserved. This is also true after $L$ spontaneous
breaking because the $L$ violating reactions are weak. The lepton number may
only be significantly violated at much lower temperatures of the order of 1
to 10 MeV when neutrino oscillations from active to sterile neutrinos become
possible. Another point is that, after $L$ breaking, the lepton number 
$L_{(\sigma )}$ carried by the scalar singlet $\sigma $ still exists but is then
associated with a coherent Majoron field. This is the subject of the next
section.


\section{Majoron field}
\label{majoron}

The Majoron equation of motion is determined by the equation
of conservation of the lepton number N\"{o}ether current. The lepton current
of the scalar field $\sigma $ with lepton number $L_{\sigma }=-2$ is 
\begin{equation}
J_{\sigma }^{\mu }=L_{\sigma }\,i\,\langle \sigma ^{*}\,\nabla ^{\mu }\sigma
-\sigma \,\nabla ^{\mu }\sigma ^{*}\rangle \;.  \label{jmusigma}
\end{equation}
At the classical level the total lepton number of charged leptons, neutrinos
and scalar $\sigma $ is conserved but electroweak instanton effects break $L$
explicitly. $B-L$ remains conserved and its equation of conservation reads
as 
\begin{equation}
\nabla _{\mu }J_{\sigma }^{\mu }+\nabla _{\mu }J_{f}^{\mu }=0\;,
\label{jconservation}
\end{equation}
where $J_{f}^{\mu }$ is the $L-B$ current of all the other particles,
 in our case leptons and quarks: 
\begin{equation}
J_{f}^{\mu }=-\sum (B_{f}-L_{f})\,(n_{f}-n_{\bar{f}})\,v^{\mu }\;,
\label{jmuf}
\end{equation}
where $n_{f}$ and $n_{\bar{f}}$ are the particle and antiparticle densities
and $v^{\mu }=(1,\mathbf{v})$ the macroscopic velocity vector.

Before spontaneous breaking of the lepton number the $\sigma $ current is
related to the $\sigma $ particle asymmetry, $J_{\sigma }^{\mu }=L_{\sigma
}\,(n_{\sigma }-n_{\bar{\sigma}})\,v^{\mu }$, but after lepton symmetry
breaking, the mass eigenstates are no longer the complex field $\sigma $ but
rather the massive Higgs particle $\rho $ and massless Majoron boson $
\varphi $. They relate to each other as 
\begin{equation}
\sigma =\frac{1}{\sqrt{2}}(v_{\sigma }+\rho )\,
\exp (-i\,\varphi /v_{\sigma })  \label{sigma}
\end{equation}
and the lepton current is expressed as 
\begin{equation}
J_{\sigma }^{\mu }=L_{\sigma }v_{\sigma }\,\langle \left( 1+\frac{\rho }{
v_{\sigma }}\right) ^{2}\nabla ^{\mu }\varphi \rangle \;.  \label{jmusigmab}
\end{equation}
As emphasized in references~\cite{bento98,gelmini83}, after symmetry
breaking the global symmetry is realized as an invariance under translations
of the Majoron field whose equation of motion is determined by the still
valid Eq.\ (\ref{jconservation}) of lepton number conservation.

After lepton breaking the current $J_{\sigma }^{\mu }$ can be realized only
through a coherent Majoron field $\varphi $; in other words, the expectation
value of $\sigma $ has a variable phase: 
\begin{equation}
\langle \sigma \rangle =\frac{v_{\sigma }}{\sqrt{2}}\exp (-i\,\varphi
/v_{\sigma })\;.
\end{equation}
We obtain for the current 
\begin{equation}
J_{\sigma }^{\mu }=L_{\sigma }v_{\sigma }\,
\left( 1+ \frac{ \langle \rho^{2}\rangle}{v_{\sigma }^{2}}  \right)
\nabla ^{\mu }\varphi \;,  \label{jmusigmac}
\end{equation}
where the term $\langle \rho ^{2}\rangle $\ is an average over quantum
fluctuations. i.e., the thermal bath of massive Higgs particles $\rho $. 
When the temperature of 
the $\rho $ bosons, lower than the photon temperature when the
number of relativistic degrees of freedom drops down one order of magnitude,
is much smaller than the breaking scale $v_{\sigma }$, the $\langle \rho
^{2}\rangle $ term can be neglected. Then, $J_{\sigma }^{\mu }=L_{\sigma
}v_{\sigma }\,\nabla ^{\mu }\varphi $. In a homogeneous and isotropic
Universe $\varphi $ depends only on time and 
\begin{equation}
J_{\sigma }^{0}=L_{\sigma }v_{\sigma }\dot{\varphi}=F_{L}\dot{\varphi}\;.
\label{j0sigma}
\end{equation}
The value of $\dot{\varphi}$ is subject to the equation of conservation (\ref
{jconservation}). Integrating over space, the $\sigma $ lepton charge $
L_{(\sigma )}=\int dV\,J_{\sigma }^{0}$ is determined at a given time by its
initial value and the variation of the $B-L$ number carried by leptons and
quarks: 
\begin{equation}
L_{(\sigma )}(t)=L_{(\sigma )}(t_{i})+(B-L)_{f}(t)-(B-L)_{f}(t_{i})\;.
\label{lsigmab}
\end{equation}
The initial value of $L_{(\sigma )}$ is the lepton charge carried by the
complex bosons $\sigma $ before spontaneous lepton breaking. 
This value is
proportional to the initial $B-L$ or to the present baryon number,
as Eq.\ (\ref{lsigma}) shows for the particular model we worked out. 
On the other hand, $B-L$ is possibly violated by neutrino oscillations only at
very low temperatures when $B$ is conserved. As a result, $L_{(\sigma
)}(t)=-A_{\sigma }B-\Delta L_{f}$ and the Majoron time derivative is
obtained from Eq.~(\ref{j0sigma}) as 
\begin{equation}
\dot{\varphi}=-\frac{n_{\gamma }}{F_{L}}(A_{\sigma }\hat{B}+\Delta \hat{L})
\label{phidot}
\end{equation}
in terms of the baryon number and lepton number variation per photon 
\begin{equation}
\hat{B}=\frac{B}{N_{\gamma }}\;,\qquad \Delta \hat{L}=\frac{\Delta L_{f}}{
N_{\gamma }}\;.  \label{bhat}
\end{equation}
Notice that $L_{f}$ and $\Delta \hat{L}$ count only the fermion particles,
charged leptons and neutrinos, but not the $\sigma $ field.


\section{Neutrino oscillations}
\label{oscillations}

Neutrino oscillations~\cite{kuo89} are governed by the neutrino
masses and mixing angles and interactions with the background 
medium~\cite{wolf78}, which
in turn depend on the temperature and particle number asymmetries. As far
as standard model weak interactions are concerned the electron, proton, and
neutron asymmetries, closely related to the baryon asymmetry, are too small
to play a role in the oscillation of active neutrinos into sterile
neutrinos. However, the neutrino asymmetries can in principle be much larger
than the baryon asymmetry. Normalizing them to the photon density as 
\begin{equation}
\hat{L}_{a}=\frac{N_{\nu _{a}}-N_{\bar{\nu}_{a}}}{N_{\gamma }}\;,  \label{la}
\end{equation}
the potential of the flavor $\nu _{a}=\nu _{e},\nu _{\mu },\nu _{\tau }$
induced by electroweak interactions is given at low temperatures ($T<m_{\mu
}\ll M_{W}$) by~\cite{notzold88} 
\begin{equation}
V_{\mathrm{EW}}=\pm \sqrt{2}G_{F}n_{\gamma }(\hat{L}_{a}+\hat{L}_{e}+\hat{L}
_{\mu }+\hat{L}_{\tau }\mp A_{a}T^{2}M_{W}^{-2})\;,  \label{vew}
\end{equation}
where $A_{e}=55$ and $A_{\mu ,\tau }=15.3$ (the electron and nucleon
asymmetries are neglected). The upper sign holds for neutrinos and the lower
sign for antineutrinos. The sterile neutrino has no
standard model potential by definition.

In the case of interest, $\delta m^{2}\sim 3\times 10^{-3}$\textrm{\ eV}$
^{2} $, the thermal contribution proportional to $n_{\gamma }T^{2}$\
prevents the oscillation into a sterile neutrino at temperatures above $\sim
10$\textrm{\ MeV}. At smaller temperatures that term becomes ineffective and
the active neutrino oscillates into the sterile flavor as in vacuum,
violating the bounds on the number of light degrees of freedom at BBN~\cite
{BBN}. Foot and Volkas~\cite{foot95} pointed out that if there is an initial
asymmetry $\hat{L }^{a}=\hat{L}_{a}+\hat{L}_{e}+\hat{L}_{\mu }+\hat{L}_{\tau
}$ larger than $7\times 10^{-5}$ and $\delta m^{2}\lesssim 10^{-2}$\textrm{
\ eV}$^{2}$, the active neutrino $\nu _{a}$ cannot significantly oscillate
into the sterile $\nu _{s}$ and the initial neutrino asymmetries are
preserved until the active neutrino $\nu _{a}$ decouples or, in the case of $
\nu _{e}$, until the protons and neutrons stop being in equilibrium. In
these conditions the BBN bounds on the extra light degrees of freedom are
satisfied. These straightforward considerations have one price, which is the
assumption of an initial neutrino asymmetry five orders of magnitude larger
than the baryon asymmetry.~\footnote{
In ref.\ \cite{foot97} Foot and Volkas explored the case where $\nu _{\tau
}\rightarrow \nu _{s}$ oscillations with $-\delta m^{2}\gtrsim 10$\textrm{\
eV}$^{2}$ and $\sin ^{2}2\theta \lesssim 10^{-5}$ create a lepton asymmetry
large enough to block $\nu _{\mu }\rightarrow \nu _{s}$ oscillations with
ANP parameters. In any case, this cannot be a generic situation and requires
some agreement in the parameter space -- 
the masses and mixing of all neutrino species.}

The situation changes if there is a Majoron field. Majorons, like any
Nambu-Goldstone boson have only derivative couplings. As a result, a
coherent Majoron field produces neutrino potentials proportional to its
gradient~\cite{bento98}. If $\Lambda $ is the spontaneously broken lepton
number, in general, any combination of partial lepton numbers, and $\Lambda
_{a}$ the quantum number of the flavor $\nu _{a}$, a Majoron field $\varphi 
$ produces the potential 
\begin{equation}
V_{\Lambda }=-\frac{1}{F_{\Lambda }}\Lambda _{a}v^{\mu }\partial _{\mu
}\varphi   \label{vlambda}
\end{equation}
for the neutrino  $\nu _{a}$
and the symmetric one for the antineutrino $\bar{\nu}_{a}$, where $v^{\mu }=(1, 
\mathbf{v})$\ is the neutrino four-velocity ($|\mathbf{v}|=1$ in leading
order). In the present case the Majoron is associated with the total lepton
number $L$ and the Majoron field is a uniform field in the Early Universe
given by Eq.~(\ref{phidot}). 
Hence, it induces the potentials\footnote{
They should not be confused with the potentials induced by a thermal bath of
Majoron particles, proportional to the neutrino masses, which
could be significant for rather large neutrino masses like
17 keV as considered in ref.\ \cite{babu92}.}
\begin{equation}
V_{L}=F_{L}^{-2}\, n_{\gamma }L_{a}(A_{\sigma }\hat{B}+\Delta \hat{L})\;,
\label{vl}
\end{equation}
where $A_{\sigma }$ is a model dependent coefficient of order 1; $A_{\sigma }\simeq 3.2$ in the case we are considering.
The quantum numbers are $L_{a}=+1$ ($-1$) for an active neutrino
(antineutrino) and $L_{a}=-3$ ($+3$) for the sterile neutrino $\nu _{s}$ ($
\bar{\nu}_{s}$).

The variation of the lepton number in the neutrino sector can be caused only
by oscillations into the sterile neutrino because this is the only one with
lepton number different from $\pm 1$ and the types of oscillations we are
considering conserve chirality. The oscillations $\nu _{a}\leftrightarrow
\nu _{s}$ ($\bar{\nu}_{a}\leftrightarrow \bar{\nu}_{s}$) produce a lepton
number variation $\Delta L=\Delta N_{a}-3\Delta N_{s}=-4\Delta N_{s}$ ($
\Delta L=4\Delta N_{\bar{s}}$); hence, 
\begin{equation}
\Delta \hat{L}=-4\frac{\Delta N_{s}-\Delta N_{\bar{s}}}{N_{\gamma }}\;.
\label{deltalhat}
\end{equation}
Combining Eqs.~(\ref{vew}) and (\ref{vl}), the difference between active and
sterile neutrino potentials is 
\begin{equation}
V_{a}-V_{s}=\sqrt{2}G_{F}n_{\gamma }(\pm \hat{L}^{a}-A_{a}T^{2}M_{W}^{-2})
\pm 4F_{L}^{-2}n_{\gamma }(A_{\sigma }\hat{B}+\Delta \hat{L})\;,
\label{va-vs}
\end{equation}
where $\hat{L}^{a}=\hat{L}_{a}+\hat{L}_{e}+\hat{L}_{\mu }+\hat{L}_{\tau }$
and the lower signs apply to antineutrinos. It is now clear that if, in the
case of standard weak interactions, an asymmetry $\hat{L}^{a}>7\times 10^{-5}$
is enough to block the $\nu _{a}\leftrightarrow \nu _{s}$, $\bar{\nu}
_{a}\leftrightarrow \bar{\nu}_{s}$ oscillations then, in the
presence of a Majoron field,
the known baryon asymmetry $\hat{B}=(4-7)\times 10^{-10}$ can do the same if
the scale of lepton number breaking $F_{L}$ obeys the condition 
\begin{equation}
F_{L}^{2}<\frac{4A_{\sigma }\hat{B}}{7\sqrt{2}G_{F}}\times 10^{5}=5-9 
\mathrm{\ \ GeV}^{2}\;.  \label{fl2}
\end{equation}

Such a low scale of lepton number  breaking is perfectly consistent with the existing
bounds for this kind of singlet Majoron 
model~\cite{chikashige81,vall82,bento98} including the astrophysical 
bounds~\cite{SN}.
The reason is that in scattering processes  the Majorons couple
primarily to neutrinos\ with strengths proportional
to the neutrino masses $g\sim m_{\nu }/F_{L}$, 
which are therefore negligibly small for the assumed
neutrino mass spectrum even if $F_{L}\sim 1\mathrm{\ GeV}$.

\section{conclusions}  \label{conclusions}

It is well known that an explanation of the  atmospheric neutrino 
problem~\cite{skatm98} in terms of oscillations of the muon neutrino into a, at least in part, sterile neutrino is in contradiction with the BBN limits on the number of extra light degrees of freedom~\cite{BBN}. 
Indeed, for such a mass gap,
$\delta m^{2}\sim 3\times 10^{-3}$ eV$^{2}$, in view of the large mixing angle,
the sterile neutrino would come into equilibrium with the
particle thermal bath via $\nu _{\mu }-\nu _{s}$ 
oscillations~\cite{dolgov81}.

In this paper we formulate a mechanism capable of blocking the active-sterile neutrino oscillations that operates in the framework of theories where the lepton number is spontaneously broken.
It has been shown~\cite{bento98} that a generic feature of these theories is the production of coherent, long-range Majoron fields.
While they can appear in stars as a result of neutrino oscillations or any other lepton number violating large scale process, we found that in the Early Universe a cosmological Majoron field emerges also as a result of a primordial lepton asymmetry carried by the scalar particles and complex scalar field ($\sigma$) whose expectation value spontaneously breaks the lepton number.
The Majoron field amplitude is thus naturally proportional to the lepton number
 of the Universe, and baryon number as well,
 due to the $L$ and $B$ violating sphaleron processes.

The leptons, and neutrinos in particular, interact with the derivatives of the Majoron field, which gives rise to new neutrino potentials that are relevant for the oscillation phenomena.
The potentials are inversely proportional to the second power of the scale of lepton number symmetry breaking ($\langle \sigma \rangle$) but this scale can be much smaller than the electroweak breaking scale.
For a lepton number breaking scale of the order of 1 GeV, a Majoron field associated with a lepton number asymmetry of the same order of magnitude as the baryon asymmetry can block the active-sterile neutrino oscillations 
at temperatures in the MeV range for 
a neutrino mass gap 
$\delta m^{2}\sim 3\times 10^{-3}$ eV$^{2}$.

This provides an interesting way to block the sterile neutrino oscillations  as it does not require the extraordinarly high lepton asymmetries,
 $\sim 10^{-5}$, that are necessary~\cite{foot95} in the framework of the standard model weak interactions.
In fact, the standard model neutrino potentials that are proportional to the background lepton asymmetries are suppressed by the Fermi constant.
In contrast, the potentials due to a Majoron field vary as the inverse square lepton number symmetry breaking scale. For a 1 GeV energy scale one immediately obtains the five orders of magnitude increase factor that brings $10^{-5}$ down to the baryon number asymmetry $B \sim 10^{-10}$.



\section*{Acknowledgments}

We acknowledge  Funda\c c\~ao para a  Ci\^encia e Tecnologia (FCT) 
for the grant CERN/P/FIS/40129/2000.  
The work of Z. B. was partially supported by the MURST research 
grant ``Astroparticle Physics".


\begin{thebibliography}{99}

\bibitem{reviews}  E. Kh. Akhmedov, 
in proceedings of the NATO ASI 
\emph{``Recent Developments in Particle Physics and Cosmology"}, 
Cascais, Portugal, June 26 - July 7, 2000, 
edited by G. C. Branco, Q. Chasi and J. I. Silva-Marcos
(Kluwer Academic, Dordrecht, in press)
 (hep-ph/0011353);
Lectures given at the ICTP \emph{``Summer School in Particle Physics''}, 
Trieste, June 7 - July 9, 1999
 (hep-ph/0001264);
S. M. Bilenky, C. Giunti and W. Grimus, 
{Prog. Part. Nucl. Phys.} {\textbf{43}}, 1 (1999).

\bibitem{sterile} For various models of sterile neutrinos, see e.g.,
A. Yu. Smirnov and J. W. F. Valle, \npbb{375}{1992}{649}; 
E. Kh. Akhmedov, Z. G. Berezhiani and G. Senjanovi\'{c}, \prll{69}{1992}{3013};
E. Kh. Akhmedov, Z. G. Berezhiani, G. Senjanovi\'{c} and Z. Tao, 
\prdd{47}{1993}{3245}; 
J. T. Peltoniemi, D. Tomassini and J. W. F. Valle, 
\plbb{298}{1993}{383};
J. T. Peltoniemi and J. W. F. Valle, \npbb{406}{1993}{409};
D. O. Caldwell and R. N. Mohapatra, \prdd{48}{1993}{3259};
E. Ma and P. Roy, Phys. Rev. D \textbf{52}, R4780 (1995); 
R. Foot and R. R. Volkas, Phys. Rev. D \textbf{52}, 6595 (1995); 
Z. G. Berezhiani and R. N. Mohapatra, Phys. Rev. D \textbf{52}, 6607 (1995); 
K. Benakli and A. Yu. Smirnov, Phys. Rev. Lett. \textbf{79}, 4314 (1997); 
G. Dvali and Y. Nir, J. High Energy Phys. \textbf{9810}, 014 (1998).  

\bibitem{skatm98}  Y. Fukuda \emph{et~al.}, Super-Kamiokande
Collaboration, {Phys. Lett. B} {\textbf{436}}, 33 (1998); {Phys. Rev. Lett.} 
{\textbf{81}}, 1562 (1998).

\bibitem{skatm00}  S. Fukuda \emph{et~al.}, Super-Kamiokande
Collaboration, {Phys. Rev. Lett.} {\textbf{85}}, 3999 (2000).

\bibitem{foot00}  R. Foot, {Phys. Lett. B} {\textbf{496}}, 169 (2000); G. L.
Fogli, E. Lisi and A. Marrone, {Phys. Rev. D} {\textbf{63}}, 053008 (2001);
{Phys. Rev. D} {\textbf{64}}, 093005 (2001).

\bibitem{BBN}  S. Burles, K. M. Nollett, J. W. Truran and M. S. Turner, Phys.
Rev. Lett. {\textbf{82}}, 4176 (1999); K. A. Olive, G. Steigman and T. P.
Walker, Phys. Rep. {\textbf{333-334}}, 389 (2000).

\bibitem{dolgov81}  A. Dolgov, {Sov. J. Nucl. Phys.} {\textbf{33}}, 700
(1981); 
R. Barbieri and A. Dolgov, {Phys. Lett. B} {\textbf{237}}, 440 (1990); 
{Nucl. Phys. B} {\textbf{349}}, 743 (1991); 
K. Kainulainen, {Phys. Lett. B} {\textbf{244}}, 191 (1990); 
K. Enqvist, K. Kainulainen and M. Thomson, 
{Nucl. Phys. B} {\textbf{373}}, 498 (1992); 
J. M. Cline, {Phys. Rev. Lett.} {\textbf{68}}, 3137 (1992); 
X. Shi, D. N. Schramm and B. D. Fields, 
{Phys. Rev. D} {\textbf{48}}, 2563 (1993).

\bibitem{kirilova97} D. P. Kirilova and M. V. Chizhov,
Phys. Lett. B \textbf{393}, 375 (1997);
Nucl. Phys. B \textbf{591}, 457 (2000);
astro-ph/0108341.

\bibitem{foot95}  R. Foot and R. R. Volkas, Phys. Rev. Lett. {\textbf{75}},
4350 (1995).

\bibitem{fukujita86}  M. Fukujita and T. Yanagida, 
Phys. Lett. B {\textbf{174}}, 45 (1986);

\bibitem{buchmuller99}  W. Buchm\"{u}ller and M. Pl\"{u}macher, 
Phys. Rep. {\textbf{320}}, 329 (1999) and Int. J. Mod. Phys. A {\textbf{15}}, 5047
(2000).

\bibitem{kuzmin85}  V. A. Kuzmin, V. A. Rubakov and M. E. Shaposhnikov,
Phys. Lett. {\textbf{155B}}, 36 (1985).

\bibitem{khlebnikov88}  S. Yu. Khlebnikov and M. E. Shaposhnikov, Nucl.
Phys. B \textbf{308}, 885 (1988).

\bibitem{harvey90}  J. A. Harvey and M. S. Turner, Phys. Rev. D {\textbf{42}}
, 3344 (1990).

\bibitem{foot96}  R. Foot, M. J. Thomson and R. R. Volkas, 
{Phys. Rev. D} {\textbf{53}}, R5349 (1996); 
X. Shi, Phys. Rev. D {\textbf{54}}, 2753 (1996);
R. Foot and R. R. Volkas, {Phys. Rev. D} {\textbf{55}}, 5147 (1997); 
R. Foot and R. R. Volkas, {Phys. Rev. D} {\textbf{56}}, 6653 (1997); 
{Phys. Rev. D} {\textbf{59}}, 029901(E) (1999); 
X. Shi and G. M. Fuller, {Phys. Rev. D} {\textbf{59}}, 063006 (1999); 
R. Foot, {Astropart. Phys.} {\textbf{10}}, 253 (1999); 
P. Di Bari, P. Lipari and M. Lusignoli, 
{Int. J. Mod. Phys. A} {\textbf{15}}, 2289 (2000); 
A. D. Dolgov, Nucl. Phys. B \textbf{610}, 411 (2001).

\bibitem{chikashige81}  Y. Chikashige, R. N. Mohapatra and R. D. Peccei, {\
Phys. Lett } {\textbf{98B}}, 265 (1981).

\bibitem{gelmini83}  G. B. Gelmini, S. Nussinov and T. Yanagida, Nucl. Phys.
B {\textbf{219}}, 31 (1983); A. A. Anselm and N. G. Uraltsev, Sov. Phys.
JETP {\textbf{57}}, 1142 (1983).

\bibitem{bento98}  L. Bento, {Phys. Rev. D} {\textbf{57}}, 583 (1998); {\
Phys. Rev. D} {\textbf{59}}, 015013 (1999); L. Bento and Z. Berezhiani, {
Phys. Rev. D} {\textbf{62}}, 055003 (2000); L. Bento, in Proceedings of the 
\emph{``X International School: Particles and Cosmology''}, Baksan Valley,
Kabardino-Balkaria, Russia,  1999, edited by E. N. Alexeev, V. A.
Matveev, Kh. S. Nirov and V. A. Rubakov (hep-ph/9908506).

\bibitem{LSND} C. Athanassopoulos \emph{et~al.}, LSND Collaboration,
 \prll{77}{1996}{3082}; 
\textbf{81}, 1774 (1998).

\bibitem{bilenky98}  S. M. Bilenky, C. Giunti and W. Grimus, 
Eur. Phys. J. C \textbf{{1}}, {247} {(1998)}; 
V. Barger, T. J. Weiler and K. Whisnant, 
\plbb{427}{1998}{97}; V. Barger, S. Pakvasa, T. J. Weiler and K. Whisnant, 
\prdd{58}{1998}{093016}; 
V. Barger, Y.-B. Dai, K. Whisnant and B.-L. Young, 
\prdd{59}{1999}{113010}.

\bibitem{bahcall01}  J. N. Bahcall, M. C. Gonzalez-Garcia and 
C. Pe\~{n}a-Garay, J. High Energy Phys. \textbf{08}, 014 (2001); 
C. Giunti, M. C. Gonzalez-Garcia, C. Pe\~{n}a-Garay,
Phys. Rev. D {\textbf{62}}, 013005 (2000).

\bibitem{barger01}  V. Barger, D. Marfatia and K. Whisnant, (hep-ph/0106207).

\bibitem{SNO01}  Q. R. Ahmad \emph{et~al.}, SNO Collaboration, Phys. Rev.
Lett. \textbf{87}, 071301 (2001).

\bibitem{CHOOZ}  M. Apollonio \emph{et~al.}, Chooz Collaboration, \plbb{420}{1998}{397}.

\bibitem{boone00}  A. O. Bazarko (MiniBooNE Collaboration), 
Nucl. Phys. Proc. Suppl. \textbf{91}, 210 (2000);
A. Para, Acta Phys. Pol. B \textbf{31}, 1313 (2000).

\bibitem{seesaw}  M. Gell-Mann, P. Ramond and R. Slansky, 
in \emph{Supergravity}, Proceedings of the Workshop, Stony Brook, N.\ Y., 1979, edited by P. van Nieuwenhuizen and D. Freedman (North-Holland, Amsterdam,
1979); 
T. Yanagida, in \emph{Proceedings of the Workshop on Unified Theories and Baryon Number in the Universe}, Tsukuda, Japan, 1979, edited by A. Sawada and A. Sugamoto
(KEK Report No. 79-18, Tsukuda, 1979); 
R. N. Mohapatra and G. Senjanovi\'{c}, Phys. Rev. Lett. \textbf{44}, 912 (1980).

\bibitem{kuo89}  T. K. Kuo and J. Pantaleone, {Rev. Mod. Phys.} 
{\textbf{61}}, 937 (1989).

\bibitem{wolf78}  L. Wolfenstein, {Phys. Rev. D} {\textbf{17}}, 2369 (1978);
S. P. Mikheyev and A. Yu. Smirnov, {Yad. Fiz.} {\textbf{42}}, 1441 (1985) 
[{Sov. J. Nucl. Phys.} {\textbf{42}}, 913 (1985)]; 
{Nuovo Cimento Soc. Ital. Fis.}, C {\textbf{9}}, 17 (1986).

\bibitem{notzold88}  D. Notzold and G. Raffelt, 
Nucl. Phys. B {\textbf{307}}, 924 (1988); 
K. Enqvist, K. Kainulainen and J. Maalampi, Phys. Lett. B {\textbf{249}},
531 (1990); Nucl. Phys. B {\textbf{349}}, 754 (1991).

\bibitem{foot97}  R. Foot and R. R. Volkas, {Phys. Rev. D} 
{\textbf{55}}, 5147 (1997); see however,
X. Shi and G. M. Fuller, {Phys. Rev. D} {\textbf{59}}, 063006 (1999).

\bibitem{babu92} K. S. Babu and I. Z. Rothstein, 
Phys. Lett. B \textbf{275}, 112 (1992).

\bibitem{vall82}  J. Schechter and J. W. F. Valle, 
Phys. Rev. D \textbf{25}, 774 (1982); 
B. Grinstein, J. Preskill and M. B. Wise, 
Phys. Lett. \textbf{159B}, 57 (1985).


\bibitem{SN}  For the experimental and astrophysical bounds on the Majoron
couplings, see J. E. Kim, Phys. Rep. \textbf{150}, 1 (1987); For the limits
imposed by SN 1987A see, e.g., K. Choi and A. Santamaria, {Phys. Rev. D} 
\textbf{42}, 293 (1990); Z. G. Berezhiani and A. Yu. Smirnov, {Phys. Lett. B} 
\textbf{220}, 279 (1989); M. Kachelriess, R. Tom\`{a}s and J. W. F. Valle,
Phys. Rev. D {\textbf{62}}, 023004 (2000); 
R. Tom\`{a}s, H. P\"{a}s, and J. W. F. Valle,
Phys. Rev. D {\textbf{64}}, 095005 (2001).

\end{thebibliography}
\end{document}